# Normal Type-2 Fuzzy Rational B-spline Curve

**Rozaimi Zakaria**

Department of Mathematics, Faculty of Science and Technology, Universiti Malaysia Terengganu, Malaysia.
rozaimi_z@yahoo.com

**Abd. Fatah Wahab**

Department of Mathematics, Faculty of Science and Technology, Universiti Malaysia Terengganu, Malaysia.
fatah@umt.edu.my

**R. U. Gobithaasan**

Department of Mathematics, Faculty of Science and Technology, Universiti Malaysia Terengganu, Malaysia.
gr@umt.edu.my

**Abstract**

In this paper, we proposed a new form of type-2 fuzzy data points(T2FDPs) that is normal type-2 data points(NT2FDPs). These brand-new forms of data were defined by using the definition of normal type-2 triangular fuzzy number(NT2TFN). Then, we applied fuzzification(alpha-cut) and type-reduction processes towards NT2FDPs after they had been redefined based on the situation of NT2FDPs. Furthermore, we redefine the defuzzification definition along with the new definitions of fuzzification process and type-reduction method to obtain crisp type-2 fuzzy solution data points. For all these processes from the defining the NT2FDPs to defuzzification of NT2FDPs, we demonstrate through curve representation by using the rational B-spline curve function as the example form modeling these NT2FDPs.

## 1 Introduction

In order to construct curves and surfaces, the collections of data are needed to make the curves and surfaces can be modeled through their existing functions [5,8,10,11,14,15,21,22,24,26,42]. These constructed curves and surfaces become the representative of the data collection to understand of the data condition and also make a conclusion of the modeling data. However, if the data become uncertain, subsequently there exist a problem in modeling the uncertainty data through curves and surface. Therefore, by the definition of type-1 fuzzy set theory, these kinds of data can be defined, which also using type-1 fuzzy number and type-1 fuzzy relation and afterwards can be modeled through curves and surface [1,4,33,35,34,38,39,37,36,41,40].

However, in dealing the problem in defining the complex uncertainty data, the existing method of type-1 fuzzy set theory(T1FST) is unable to apply in order to define them. Therefore, Zadeh comes with the new theory which known as type-2 fuzzy set theory(T2FST) [44]. By using this theory and the concept of type-2 fuzzy number(T2FN), we can define the complex uncertainty data which become type-2 fuzzy data. For real data problem, we also used T2FST, T2FN and type-2 fuzzy relation(T2FR) to define them, which become type-2 fuzzy data points(T2FDP).

Due to the new definition of defining complex uncertainty data through T2FST, we will discuss about the fuzzification process(alpha-cut operation) which is been applied first before we come to the next method in getting the brand-new type-2 fuzzy interval points from the old type-2 fuzzy interval points of T2FDPs. Then, we used the existing type-1 defuzzification method to defuzzify the new T2FDPs for obtaining crisp type-2 fuzzy solution data points.

This paper organized as follows: Section 2 discusses of some basics definitions of T2FST. Then, in Section 3, we discuss about the T2FDPs definition along with the alpha-cut operation process, type-reduction and also defuzzification methods. For Section 4, we will discuss about the development of NT2FDPs based on the definition of NT2TFN and the processes of getting crisp type-2 fuzzy solution data points before we demonstrate for all processes that had been mentioned before by using the rational B-spline curve function as the hypothetical example as in Section 5. Then, this proposed method can be summarized in Section 6.

## 2 Type-2 Fuzzy Set Theory

T2FST was introduced by Zadeh in 1975 [44] to solve the problem in defining the complex uncertainty which the problem is unable to define the existing T1FST. Here, we will define some basic theories of T2FST, which given as follows.

**Definition 2.1.** A type-2 fuzzy set(T2FS), denoted $\ddot{\ddot{A}}$, is characterized by a type-2 membership function $\mu_{\ddot{\ddot{A}}}(x,u)$, where $x \in X$ and $u \in U_x \subseteq [0,1]$ that is,

$$\ddot{\ddot{A}} = \left\{ \left( (x,u), \mu_{\ddot{\ddot{A}}}(x,u) \right) \mid \forall x \in X, \forall u \in U_x \subseteq [0,1] \right\}$$

in which, $0 \le \mu_{\tilde{\tilde{A}}}(x,u) \le 1$ [28].

**Definition 2.2.** A T2FN is broadly defined as a T2FS that has a numerical domain. An interval T2FS is defined using the following four constraints, where $\tilde{\tilde{A}}_\alpha = \{[a^\alpha, b^\alpha], [c^\alpha, d^\alpha]\}$, $\forall \alpha \in [0,1]$, $\forall a^\alpha, b^\alpha, c^\alpha, d^\alpha \in \mathbb{R}$ (Fig. 2.1) [25,45]:

1. $a^\alpha \le b^\alpha \le c^\alpha \le d^\alpha$
2. $[a^\alpha, d^\alpha]$ and $[b^\alpha, c^\alpha]$ generate a function that is convex and $[a^\alpha, d^\alpha]$ generate a function is normal.
3. $\forall \alpha_1, \alpha_2 \in [0,1] : (\alpha_2 > \alpha_1) \Rightarrow \left(\left[a^{\alpha_1}, c^{\alpha_1}\right] \supset \left[a^{\alpha_2}, c^{\alpha_2}\right],\right.$ $\left.\left[b^{\alpha_1}, d^{\alpha_1}\right] \supset \left[b^{\alpha_2}, d^{\alpha_2}\right]\right)$, for $c^{\alpha_2} \ge b^{\alpha_2}$.
4. If the maximum of the membership function generated by $[b^\alpha, c^\alpha]$ is the level $\alpha_m$, that is, $[b^{\alpha_m}, c^{\alpha_m}]$, then $\left[b^{\alpha_m}, c^{\alpha_m}\right] \subset \left[a^{\alpha=1}, d^{\alpha=1}\right]$.

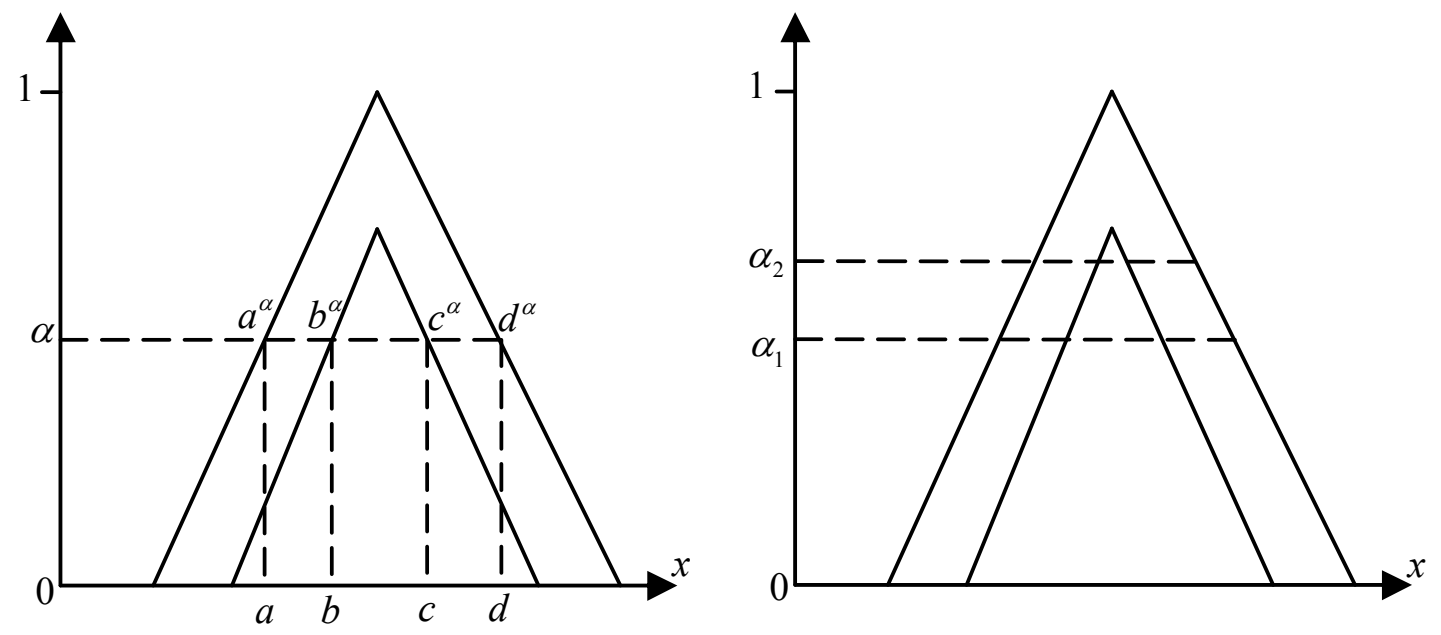

Figure 2.1. Definition of an interval T2FN.

**Definition 2.3.** Let $X, Y, U_x, V_y \subseteq R$ and

$\tilde{\tilde{A}} = \left\{\left((x,u), \mu_{\tilde{\tilde{A}}}(x,u)\right) \mid \forall x \in X, \forall u \in U_x \subseteq [0,1]\right\}$ and

$\tilde{\tilde{B}} = \left\{\left((y,v), \mu_{\tilde{\tilde{B}}}(y,v)\right) \mid \forall y \in Y, \forall v \in V_y \subseteq [0,1]\right\}$

are two T2FSs. Then, $\tilde{\tilde{R}} = \left\{\left(\left((x,u),(y,v)\right),\right.\right.$ $\left.\left.\mu_{\tilde{\tilde{R}}}\left(\mu_{\tilde{\tilde{A}}}(x,u), \mu_{\tilde{\tilde{B}}}(y,v)\right)\right) \mid \left(\forall x \in X, \forall u \in U_x\right) \times \left(\forall y \in Y, \forall u \in V_y\right) \subseteq [0,1]\right\}$ is a T2FR on $\tilde{\tilde{A}}$ and $\tilde{\tilde{B}}$ if $\mu_{\tilde{\tilde{R}}}\left(\mu_{\tilde{\tilde{A}}}(x,u), \mu_{\tilde{\tilde{B}}}(y,v)\right) \le \mu_{\tilde{\tilde{A}}}(x,u)$, $\forall \left((x,u),(y,v)\right) \in \left(\forall x \in X, \forall u \in U_x\right) \times$ $\left(\forall y \in Y, \forall v \in V_y\right)$.

# 3 Type-2 Fuzzy Data Points

After the T2FST, T2FN and T2FR had been defined, then we used them to define the complex uncertainty data which will become T2FDPs is given as follows.

**Definition 3.1.** Let $P=\{x \mid x \text{ type-2 fuzzy point}\}$ and $\overleftrightarrow{\overleftrightarrow{P}}=\{P_i \mid P_i \text{ data point}\}$ which is set of type-2 fuzzy data point with $P_i \in P \subset X$, where $X$ is a universal set and $\mu_P(P_i): P \to [0,1]$ is the membership function which defined as $\mu_P(P_i)=1$ and formulated by $\overleftrightarrow{\overleftrightarrow{P}}=\{(P_i,\mu_P(P_i)) \mid P_i \in \square\}$ . Therefore,

$$\mu_P(P_i)=\begin{cases} 0 & \text{if } P_i \notin X \\ c \in (0,1) & \text{if } P_i \tilde{\in} X \\ 1 & \text{if } P_i \in X \end{cases} \tag{3.1}$$

with $\mu_P(P_i)=\left\langle \mu_P(\overleftrightarrow{\overleftrightarrow{P}}_i^{\leftarrow}), \mu_P(P_i), \mu_P(\overleftrightarrow{\overleftrightarrow{P}}_i^{\rightarrow}) \right\rangle$ which $\mu_P(\overleftrightarrow{\overleftrightarrow{P}}_i^{\leftarrow})$ and $\mu_P(\overleftrightarrow{\overleftrightarrow{P}}_i^{\rightarrow})$ are left and right footprint of membership values with $\mu_P(\overleftrightarrow{\overleftrightarrow{P}}_i^{\leftarrow})=\left\langle \mu_P(\overleftarrow{\overleftrightarrow{P}_i^{\leftarrow}}), \mu_P(\overleftrightarrow{P}_i^{\leftarrow}), \mu_P(\overrightarrow{\overleftrightarrow{P}_i^{\leftarrow}}) \right\rangle$ where, $\mu_P(\overleftarrow{\overleftrightarrow{P}_i^{\leftarrow}})$, $\mu_P(\overleftrightarrow{P}_i^{\leftarrow})$ and $\mu_P(\overrightarrow{\overleftrightarrow{P}_i^{\leftarrow}})$ are left-left, left, right-left membership grade values and $\mu_P(\overrightarrow{\overleftrightarrow{P}_i^{\rightarrow}})$, $\mu_P(\overleftrightarrow{P}_i^{\rightarrow})$ and $\mu_P(\overleftarrow{\overleftrightarrow{P}_i^{\rightarrow}})$ are right-right, right, left-right membership grade values, which can be written as

$$\overleftrightarrow{\overleftrightarrow{P}}=\left\{\overleftrightarrow{\overleftrightarrow{P}}_i : i=0,1,2,\ldots,n\right\} \tag{3.2}$$

for every $i$, $\overleftrightarrow{\overleftrightarrow{P}}_i=\left\langle \overleftrightarrow{\overleftrightarrow{P}}_i^{\leftarrow}, P_i, \overleftrightarrow{\overleftrightarrow{P}}_i^{\rightarrow} \right\rangle$ with $\overleftrightarrow{\overleftrightarrow{P}}_i^{\leftarrow}=\left\langle \overleftarrow{\overleftrightarrow{P}_i^{\leftarrow}}, \overleftrightarrow{P}_i^{\leftarrow}, \overrightarrow{\overleftrightarrow{P}_i^{\leftarrow}} \right\rangle$ where $\overleftarrow{\overleftrightarrow{P}_i^{\leftarrow}}$, $\overleftrightarrow{P}_i^{\leftarrow}$ and $\overrightarrow{\overleftrightarrow{P}_i^{\leftarrow}}$ are left-left, left and right-left T2FDPs and $\overleftrightarrow{\overleftrightarrow{P}}_i^{\rightarrow}=\left\langle \overleftarrow{\overleftrightarrow{P}_i^{\rightarrow}}, \overleftrightarrow{P}_i^{\rightarrow}, \overrightarrow{\overleftrightarrow{P}_i^{\rightarrow}} \right\rangle$ where $\overleftarrow{\overleftrightarrow{P}_i^{\rightarrow}}$, $\overleftrightarrow{P}_i^{\rightarrow}$ and $\overrightarrow{\overleftrightarrow{P}_i^{\rightarrow}}$ are left-right, right and right-right T2FDPs respectively. This can be illustrated as in Fig. 3.1.

The illustration of T2FDP was shown in Fig. 3.1 which T1FDP becomes the primary membership function bounded by upper bound, $\left[\overleftarrow{\overleftrightarrow{P}^{\leftarrow}}, P, \overrightarrow{\overleftrightarrow{P}^{\rightarrow}}\right]$ and lower bound, $\left[\overrightarrow{\overleftrightarrow{P}^{\leftarrow}}, P, \overleftarrow{\overleftrightarrow{P}^{\rightarrow}}\right]$ respectively. The process of defining T2FDP can be shown through Fig. 3.2.

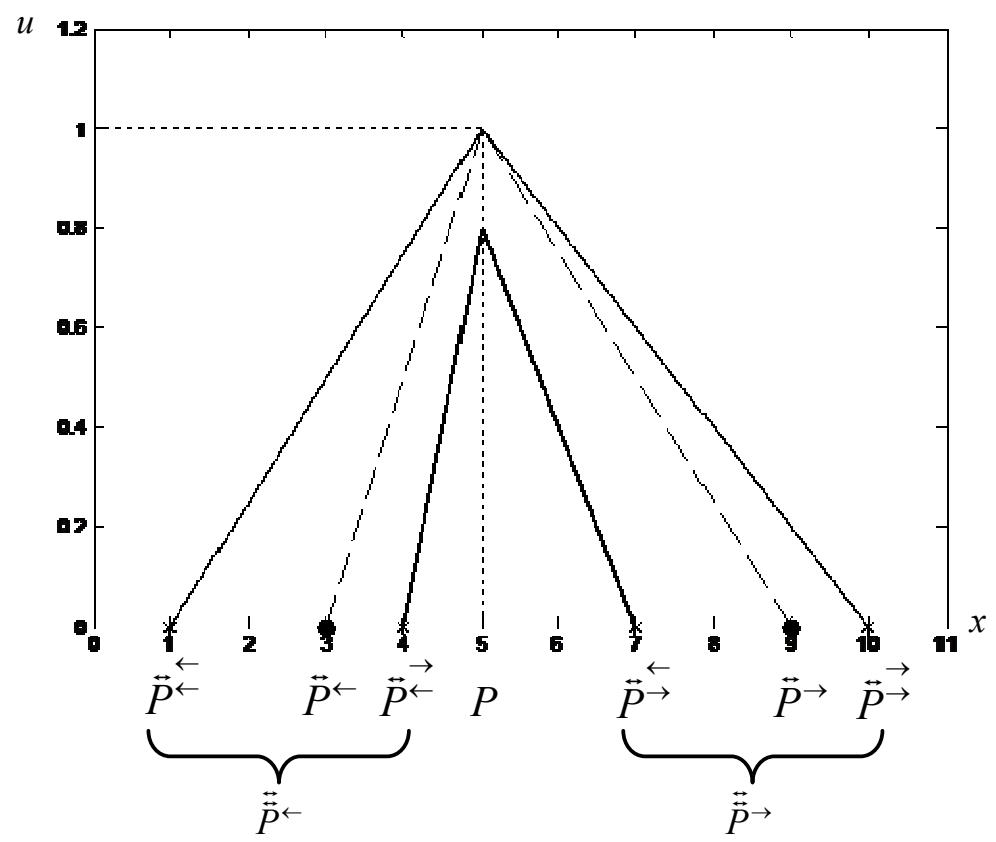

Figure 3.1. T2FDP around 5.

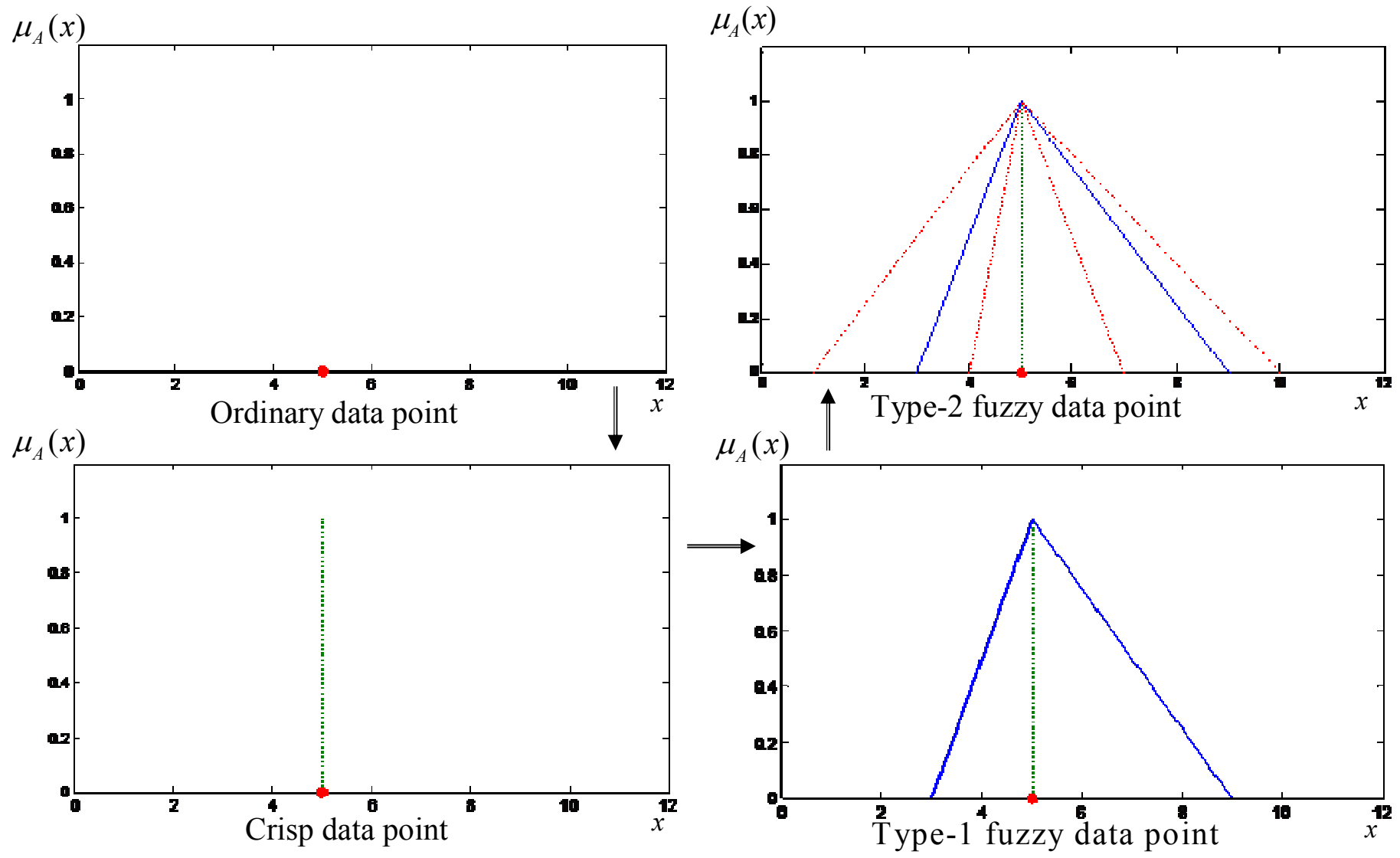

Figure 3.2. Process of defining T2FDP.

Fig. 3.2 shows that the process of defining T2FDP from the ordinary point. This T2FDP formed based on the definition of T2FN and T2FR.

## 4 Normal Type-2 Fuzzy Data Points

In this Section 4, we will discuss about the definition of NT2FDP together with its fuzzification process(alpha-cut operation), type-reduction and defuzzification methods.

**Definition 4.1.** Given that T2FDP, $\overleftrightarrow{P}$ which the height of LMF and UMF are $h\left(\overleftrightarrow{\underset{\smile}{P}}\right)$ and $h\left(\overleftrightarrow{\underset{\frown}{P}}\right)$ respectively, then T2FDP called **Normal** of T2FDP(NT2FDP) if $h\left(\overleftrightarrow{\underset{\smile}{P}}\right) < h\left(\overleftrightarrow{\underset{\frown}{P}}\right) = 1$ [20]. This Def. 4.1 can be illustrated through Fig. 4.1.

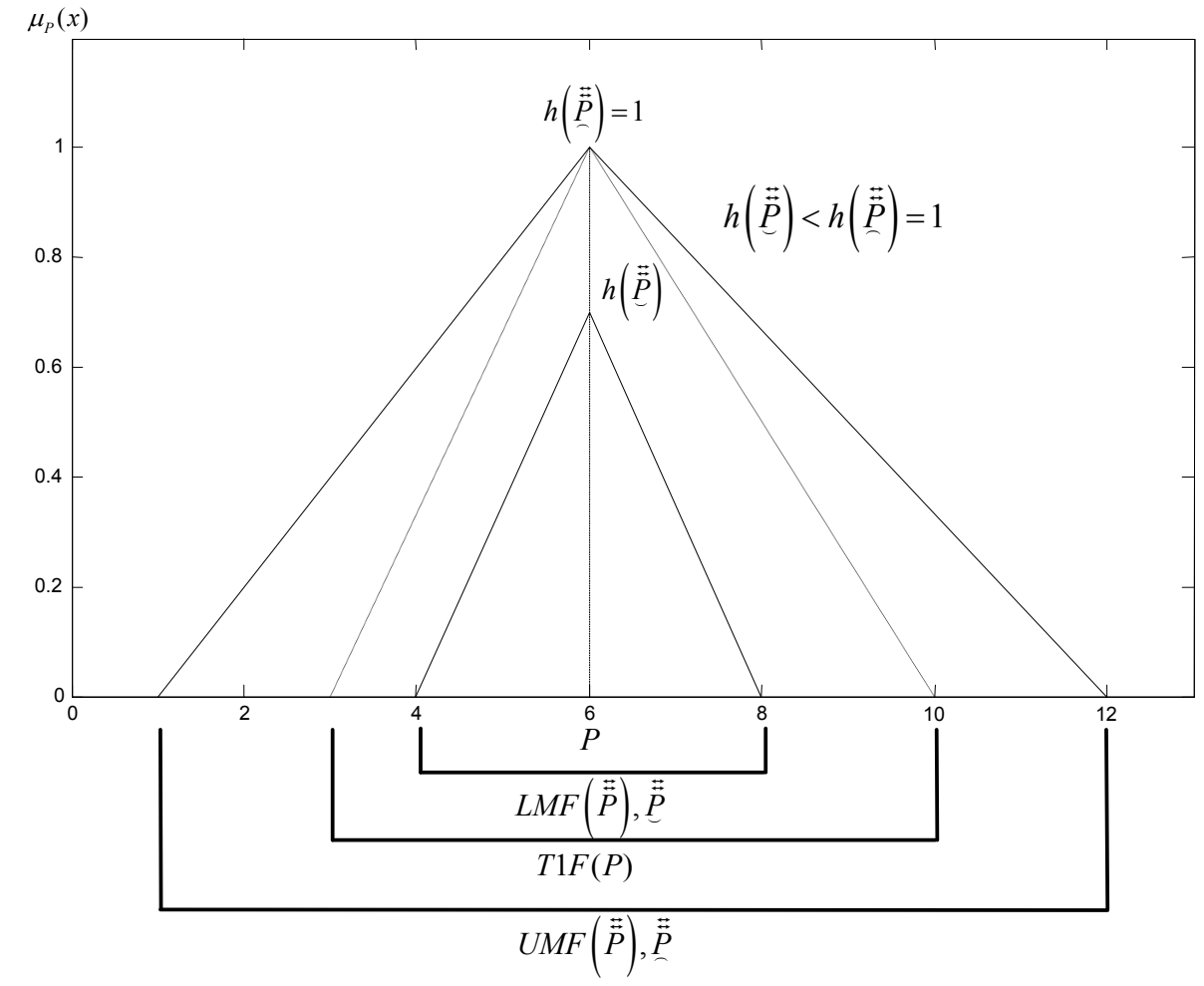

Figure 4.1. The NT2FDPs.

**Definition 4.2.** Based on Def. 3.1, let $\overleftrightarrow{P}$ be the set of T2FDPs with $\overleftrightarrow{P}_i \in \overleftrightarrow{P}$ where $i = 0,1,...,n-1$. Then ${}^{N}\overleftrightarrow{P}^{\alpha}$ is the $\alpha$-cut operation of normal T2FDPs which is given as equation as follows for $\alpha_i < \alpha_{LMF} = h\left(\overleftrightarrow{\underset{\smile}{P}}\right) < \alpha_{UMF} = h\left(\overleftrightarrow{\underset{\frown}{P}}\right)$.

$$
\begin{aligned}
{}^{N}\overleftrightarrow{P}_i^{\alpha} &= \left\langle \overleftrightarrow{P}_i^{\alpha\leftarrow}, P_i, \overleftrightarrow{P}_i^{\alpha\rightarrow} \right\rangle \\
&= \left\langle \left\langle \overleftarrow{\overleftrightarrow{P}_i^{\alpha\leftarrow}}; \overleftrightarrow{P}_i^{\alpha\leftarrow}; \overrightarrow{\overleftrightarrow{P}_i^{\alpha\leftarrow}} \right\rangle, P_i, \left\langle \overleftarrow{\overleftrightarrow{P}_i^{\alpha\rightarrow}}; \overleftrightarrow{P}_i^{\alpha\rightarrow}; \overrightarrow{\overleftrightarrow{P}_i^{\alpha\rightarrow}} \right\rangle \right\rangle \\
&= \left\langle \left[ \left( P_i - \left\langle \overleftarrow{\overleftrightarrow{P}_i^{\leftarrow}}; \overleftrightarrow{P}_i^{\leftarrow}; \overrightarrow{\overleftrightarrow{P}_i^{\leftarrow}} \right\rangle \right) \left\langle \alpha; \alpha; \frac{\alpha_{LMF}}{\alpha_{CLMF}} \right\rangle + \left\langle \overleftarrow{\overleftrightarrow{P}_i^{\leftarrow}}; \overleftrightarrow{P}_i^{\leftarrow}; \overrightarrow{\overleftrightarrow{P}_i^{\leftarrow}} \right\rangle \right], P_i, \right. \\
&\qquad \left. \left[ -\left( \left\langle \overleftarrow{\overleftrightarrow{P}_i^{\rightarrow}}; \overleftrightarrow{P}_i^{\rightarrow}; \overrightarrow{\overleftrightarrow{P}_i^{\rightarrow}} \right\rangle - P_i \right) \left\langle \frac{\alpha_{LMF}}{\alpha_{CLMF}}; \alpha; \alpha \right\rangle + \left\langle \overleftarrow{\overleftrightarrow{P}_i^{\rightarrow}}; \overleftrightarrow{P}_i^{\rightarrow}; \overrightarrow{\overleftrightarrow{P}_i^{\rightarrow}} \right\rangle \right] \right\rangle
\end{aligned}
\tag{4.1}
$$

where $\alpha_{LMF}$ and $\alpha_{CLMF}$ are alpha values of lower membership function and crisp lower membership function of normal T2FDPs respectively. This definition can be illustrated through Fig. 4.2.

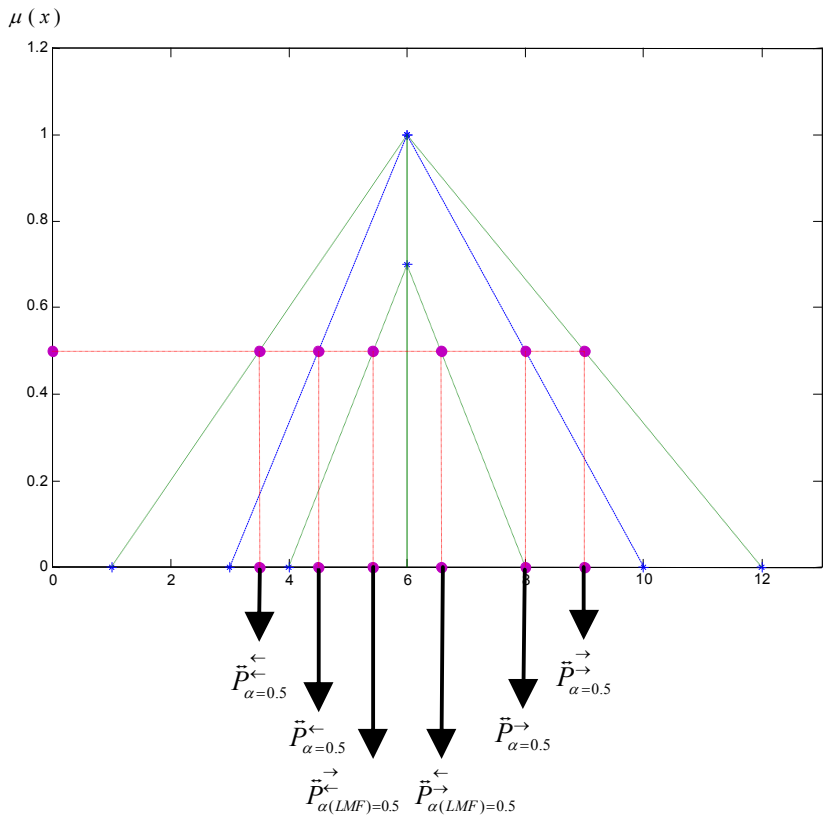

Figure 4.2. The alpha-cut operation towards normal T2FDP.

However, when $\alpha_{LMF} < \alpha_i < \alpha_{UMF}$ for $\alpha$ -cut operation of normal T2FDPs, then the Eq. 4.1 become

$$
\begin{aligned}
{}^{N}\vec{\overleftrightarrow{P}}_i^{\alpha} &= \left\langle \vec{\overleftrightarrow{P}}_i^{\alpha\leftarrow}, P_i, \vec{\overleftrightarrow{P}}_i^{\alpha\rightarrow} \right\rangle \\
&= \left\langle \left\langle \vec{\overleftarrow{P}}_i^{\alpha\leftarrow}; \vec{P}_i^{\alpha\leftarrow}; 0\right\rangle, P_i, \left\langle 0; \vec{P}_i^{\alpha\rightarrow}; \vec{\overrightarrow{P}}_i^{\alpha\rightarrow}\right\rangle\right\rangle \\
&= \Big\langle \Big[ \Big( P_i - \left\langle \vec{\overleftarrow{P}}_i^{\leftarrow}; \vec{P}_i^{\leftarrow}; 0 \right\rangle \Big) \langle \alpha;\alpha;0\rangle + \left\langle \vec{\overleftarrow{P}}_i^{\leftarrow}; \vec{P}_i^{\leftarrow};0\right\rangle \Big], P_i, \\
&\quad \Big[ -\Big( \left\langle 0; \vec{P}_i^{\rightarrow}; \vec{\overrightarrow{P}}_i^{\rightarrow}\right\rangle - P_i \Big)\langle 0;\alpha;\alpha\rangle + \left\langle 0;\vec{P}_i^{\rightarrow};\vec{\overrightarrow{P}}_i^{\rightarrow}\right\rangle \Big]\Big\rangle
\end{aligned} \tag{4.2}
$$

which can be illustrated by given this following figure.

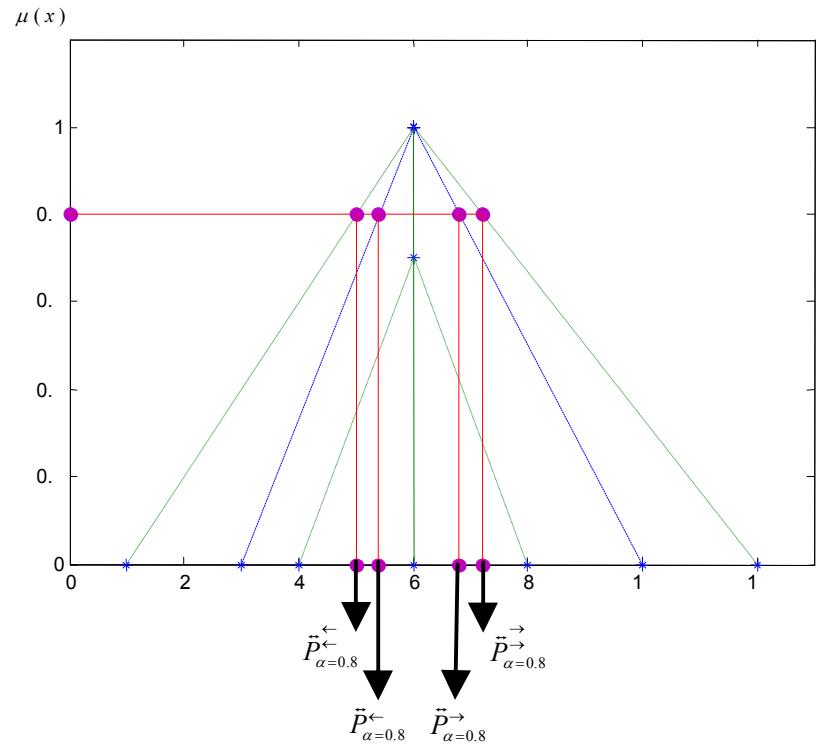

Figure 4.3. The alpha-cut operation towards NT2FDP with $\alpha_{LMF} < \alpha < \alpha_{UMF}$ .

After the alpha-cut operation against NT2FDPs had been applied, then the next step for obtaining crisp type-2 fuzzy solution data point is the type-reduction method. Type-reduction method is a method were used to reduce T2FDPs to T1FDPs after fuzzification process has been applied for allowing the defuzzification of a type-1 cases can be applied. There are many types of type-reduction method, which can be referred in [6,7,9,16-19,23,28,29,43]. However, in this paper, we construct another type-reduction method, which is known as centroid min method, which can be given via Def. 4.3.

**Definition 4.3.** Let ${}^{N}\overset{\leftrightarrow}{\overset{\leftrightarrow}{P}}_i$ be a set $(n+1)$ NT2FDPs, then type-reduction method of $\alpha$-T2FDPs(after fuzzification), ${}^{N}\overline{\overset{\leftrightarrow}{P}}_i$ is defined by

$$ {}^{N}\overline{\overset{\leftrightarrow}{P}}^{\alpha} = \left\{ \overline{\overset{\leftrightarrow}{P}}_i^{\alpha} = \left\langle \overline{\overset{\leftrightarrow}{P}}_i^{\alpha\leftarrow}, P_i, \overline{\overset{\leftrightarrow}{P}}_i^{\alpha\rightarrow} \right\rangle ; i = 0,1,\ldots,n \right\} \tag{4.3} $$

where $\overline{\overset{\leftrightarrow}{P}}_i^{\alpha\leftarrow}$ is left type-reduction of $\alpha$-cut NT2FDPs, $\overline{\overset{\leftrightarrow}{P}}_i^{\alpha\leftarrow} = \frac{1}{3} \sum_{i=0,\ldots,n} \left\langle \overset{\leftarrow}{\overset{\leftrightarrow}{P}}{}_i^{\alpha\leftarrow} + \overset{\leftrightarrow}{P}_i^{\alpha\leftarrow} + \overset{\rightarrow}{\overset{\leftrightarrow}{P}}{}_i^{\alpha\leftarrow} \right\rangle$, $P_i$ is the crisp point and $\overline{\overset{\leftrightarrow}{P}}_i^{\alpha\rightarrow}$ is right type-reduction of $\alpha$-cut NT2FDPs, $\overline{\overset{\leftrightarrow}{P}}_i^{\alpha\rightarrow} = \frac{1}{3} \sum_{i=0,\ldots,n} \left\langle \overset{\leftarrow}{\overset{\leftrightarrow}{P}}{}_i^{\alpha\rightarrow} + \overset{\leftrightarrow}{P}_i^{\alpha\rightarrow} + \overset{\rightarrow}{\overset{\leftrightarrow}{P}}{}_i^{\alpha\rightarrow} \right\rangle$ for $\alpha_i < \alpha_{LMF} < \alpha_{UMF}$.

However, same goes with the condition through Eq. 4.2, then the type-reduction NT2FDPs for $\alpha_{LMF} < \alpha_i < \alpha_{UMF}$ is given by Eq. 4.3 where $\overline{\overset{\leftrightarrow}{P}}_i^{\alpha\leftarrow}$ is left type-reduction of $\alpha$-cut NT2FDPs, $\overline{\overset{\leftrightarrow}{P}}_i^{\alpha\leftarrow} = \frac{1}{2} \sum_{i=0,\ldots,n} \left\langle \overset{\leftarrow}{\overset{\leftrightarrow}{P}}{}_i^{\alpha\leftarrow} + \overset{\leftrightarrow}{P}_i^{\alpha\leftarrow} + 0 \right\rangle$ and $\overline{\overset{\leftrightarrow}{P}}_i^{\alpha\rightarrow}$ is right type-reduction of $\alpha$-cut NT2FDPs, $\overline{\overset{\leftrightarrow}{P}}_i^{\alpha\rightarrow} = \frac{1}{2} \sum_{i=0,\ldots,n} \left\langle 0 + \overset{\leftrightarrow}{P}_i^{\alpha\rightarrow} + \overset{\rightarrow}{\overset{\leftrightarrow}{P}}{}_i^{\alpha\rightarrow} \right\rangle$. Then, to find the defuzzify of T2FDPs after type-reduction method has been applied is using the defuzzification method for T1FDPs cases, which defined in Def. 4.4.

**Definition 4.4.** Let $\alpha$-TR is the type-reduction method after $\alpha$-cut process had been applied for every T2FDPs, ${}^{N}\overline{\overset{\leftrightarrow}{P}}_i^{\alpha}$. Then ${}^{N}\overline{\overline{P}}_i^{\alpha}$ named as defuzzification NT2FDPs for ${}^{N}\overline{\overset{\leftrightarrow}{P}}_i^{\alpha}$ if for every ${}^{N}\overline{\overset{\leftrightarrow}{P}}_i^{\alpha} \in {}^{N}\overline{\overset{\leftrightarrow}{P}}^{\alpha}$,

$$ {}^{N}\overline{\overline{P}}^{\alpha} = \left\{ {}^{N}\overline{\overline{P}}_i^{\alpha} \right\} \quad \text{for} \quad i = 0,1,\ldots,n \tag{4.4} $$

where for every ${}^{N}\overline{\overline{P}}_i^{\alpha} = \frac{1}{3} \sum_{i=0} < \overline{\overset{\leftrightarrow}{P}}_i^{\alpha\leftarrow}, P_i, \overline{\overset{\leftrightarrow}{P}}_i^{\alpha\rightarrow} >$. The process in defuzzifying

NT2FDPs can be illustrated at Fig. 4.4.

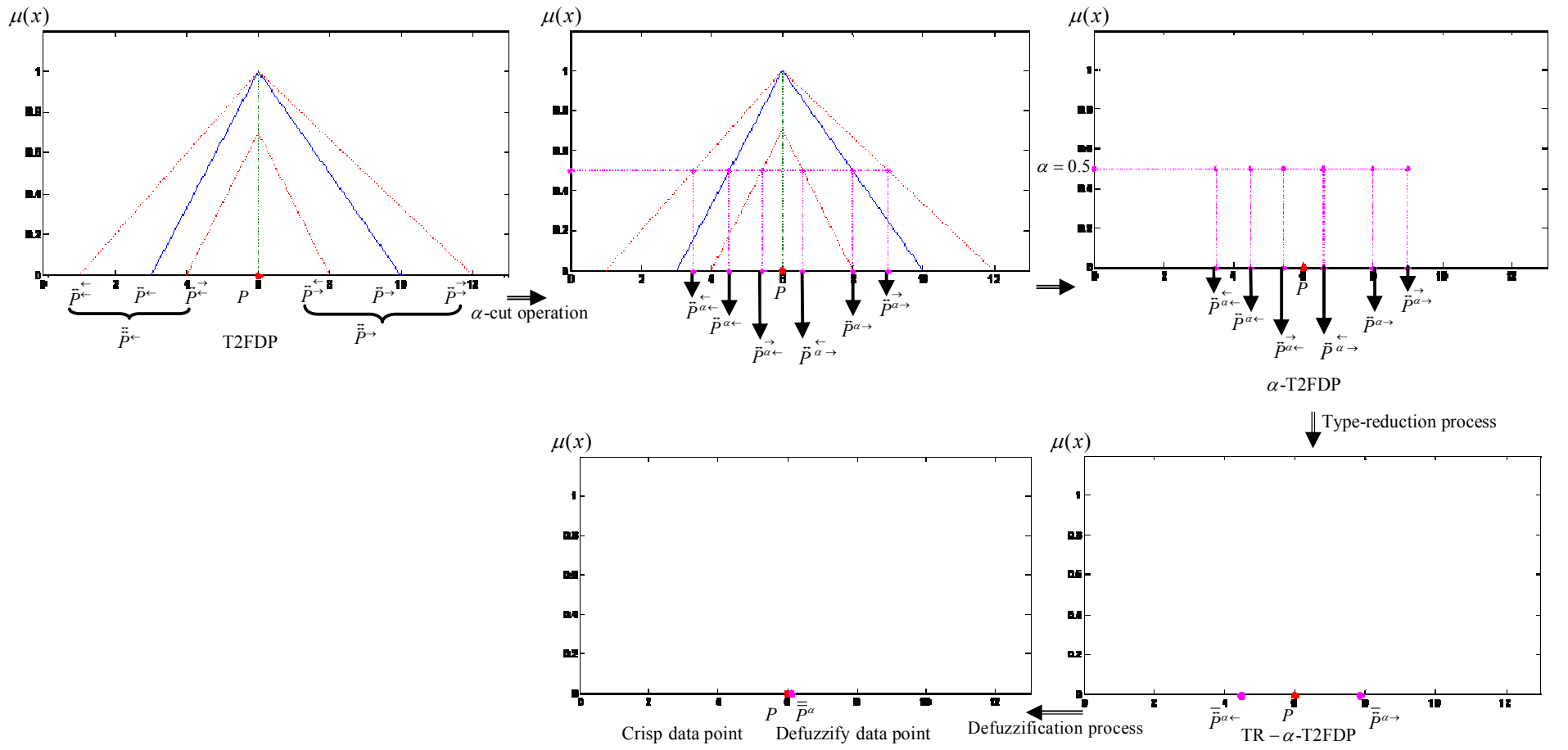

Figure 4.4. Defuzzification process of NT2FDP.

Fig. 4.4 shows that the process of defuzzying the T2FDP which gives the crisp type-2 fuzzy solution of data points as a final result in define the complex uncertainty data points.

## 5 Normal Type-2 Fuzzy Rational B-spline Curve Modeling

After we had finished in defining the complex uncertainty data points using T2FST, which gives T2FDP in NT2FDP form, then we use the rational B-spline curve function in order to model the NT2FDPs which finally known as normal type-2 fuzzy rational B-spline curve(NT2FRBsC).

NT2FRBsC will give us more comprehended in how normal T2FDPs can be modeled and influence the rational curve form. Besides that, we can use the definition of NT2FDP in defining the real complex uncertainty data which we want to model by using the existing curves and surfaces function. Therefore, as the hypothetical example of modeling the complex uncertainty data points, the process of modeling for NT2FDPs using Rational B-spline curve function [12,13,27,30-32,42] can be given in Def. 5.1 as follows.

**Definition 5.1.** Let ${}^{N}\overleftrightarrow{\overleftrightarrow{RBsC}}_{k,n}(t)$ denote a normal type-2 fuzzy rational B-spline of order $k$ (degree, $k$-1), where $k \leq n$ . Let $w_i$ for $i = 1,...,n$ be the $n$ weights corresponding to homogeneous normal type-2 fuzzy data/control points $\left\{ {}^{N}\overleftrightarrow{\overleftrightarrow{P}}_1, {}^{N}\overleftrightarrow{\overleftrightarrow{P}}_2, ..., {}^{N}\overleftrightarrow{\overleftrightarrow{P}}_n \right\}$. Therefore, the NT2FRBsC becomes:

$$ {}^{N}\overleftrightarrow{\overleftrightarrow{RBsC}}_{k,n}(t) = \frac{\sum_{i=1}^{n} w_i N_i^k(t)\, {}^{N}\overleftrightarrow{\overleftrightarrow{P}}_i}{\sum_{r=1}^{n} w_r N_r^k(t)} \tag{5.1} $$

where $N$ is the B-spline basis function.

After we defined the $\overleftrightarrow{^{N}RBsC_{k,n}}(t)$ by using the definition of T2FST and the other's relation definitions, then we modeled the $\overleftrightarrow{^{N}RBsC_{k,n}}(t)$ together with the alpha-cut operation, type-reduction process and defuzzification method in Alg. 1.1 as follows.

**Algorithm 1.1.** Algorithm of fuzzification, type-reduction and defuzzification processes of NT2FRBsC for $\alpha_{LMF} < \alpha_i < \alpha_{UMF}$ case.

**Step 1:** Define four NT2FCPs, ${}^{N}\overset{\leftrightarrow}{\overset{\leftrightarrow}{P}}_{i=1,\ldots,4}$. Then, the NT2FRBsC equation can be given by Eq. 5.1 as illustrated by Fig. 5.1.

$$\overleftrightarrow{^{N}RBsC_{k,n}}(t) = \frac{\sum_{i=1}^{n} w_i N_i^k(t)\, {}^{N}\overset{\leftrightarrow}{\overset{\leftrightarrow}{P}}_i}{\sum_{r=1}^{n} w_r N_r^k(t)}, \quad k = 3;\ n = 4;\ i = r = 1,\ldots,4.$$

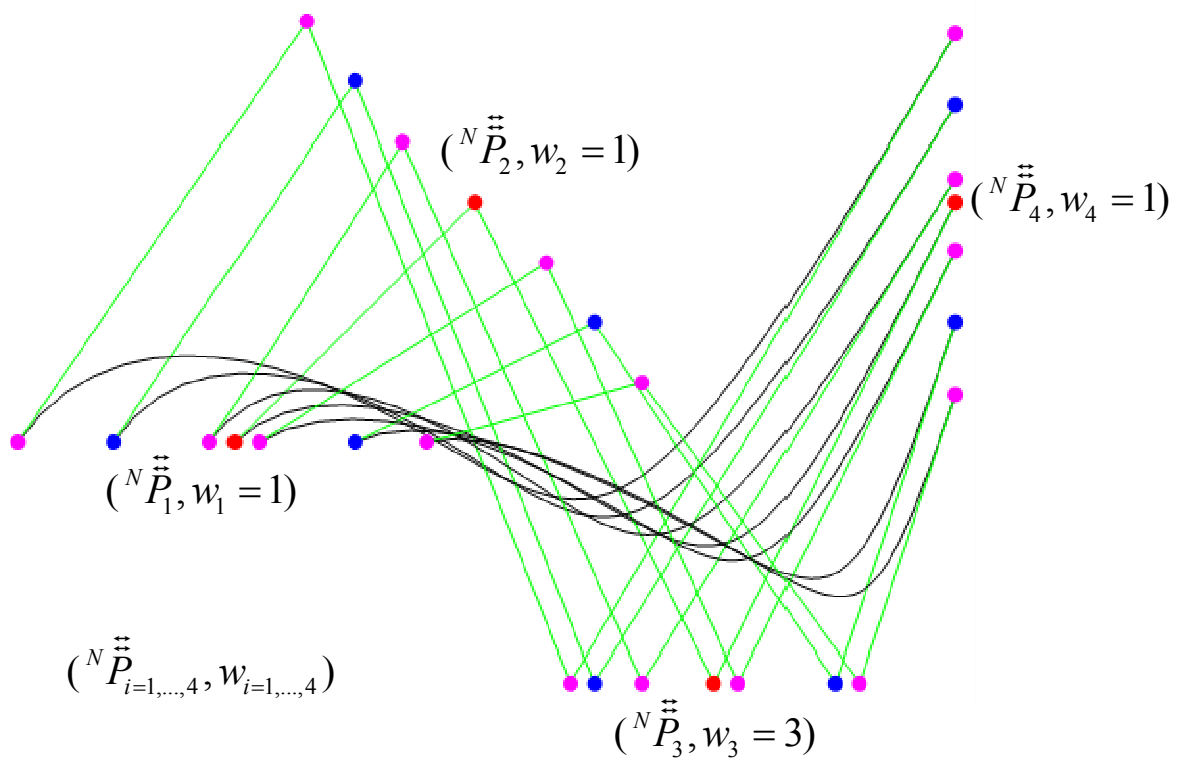

Figure 5.1. The modeling of NT2FRBsC with $w_{i=1,\ldots,4} = (1,1,3,1)$.

**Step 2:** Fuzzification(alpha-cut operation).

For left($x$-axis)/lower($y$-axis),

$$\left({}^{N}\overset{\leftrightarrow}{\overset{\leftrightarrow}{P}}{}^{\leftarrow}_{i=1,\ldots,4} - P_{i=1,\ldots,4}\right)\alpha_{0.8} + {}^{N}\overset{\leftrightarrow}{\overset{\leftrightarrow}{P}}{}^{\leftarrow}_{i=1,\ldots,4} \text{ where } {}^{N}\overset{\leftrightarrow}{\overset{\leftrightarrow}{P}}{}^{\leftarrow}_{i=1,\ldots,4} = \left\langle \overleftarrow{\vec{P}_i^{\leftarrow}}, \vec{P}_i^{\leftarrow}, 0 \right\rangle_{i=1,\ldots,4}. \tag{5.2}$$

For right($x$-axis)/upper($y$-axis),

$$\left({}^{N}\overset{\leftrightarrow}{\overset{\leftrightarrow}{P}}{}^{\rightarrow}_{i=1,\ldots,4} - P_{i=1,\ldots,4}\right)\alpha_{0.8} + {}^{N}\overset{\leftrightarrow}{\overset{\leftrightarrow}{P}}{}^{\rightarrow}_{i=1,\ldots,4} \text{ where } {}^{N}\overset{\leftrightarrow}{\overset{\leftrightarrow}{P}}{}^{\rightarrow}_{i=1,\ldots,4} = \left\langle 0, \vec{P}_i^{\rightarrow}, \overrightarrow{\vec{P}_i^{\rightarrow}} \right\rangle_{i=1,\ldots,4}. \tag{5.3}$$

For alpha-cut of diagonal points, used Eq. 5.2 if ${}^{N}\overset{\leftrightarrow}{\overset{\leftrightarrow}{P}}{}^{\leftarrow}(x,y) < P(x,y)$ or ${}^{N}\overset{\leftrightarrow}{\overset{\leftrightarrow}{P}}{}^{\rightarrow}(x,y) < P(x,y)$ and Eq. 5.3 if $P(x,y) < {}^{N}\overset{\leftrightarrow}{\overset{\leftrightarrow}{P}}{}^{\leftarrow}(x,y)$ or

$P(x,y) < {}^{N}\overset{\leftrightarrow}{\vec{P}}^{\rightarrow}(x,y)$.

Then, the brand-new NT2FDPs( $\alpha$ -NT2FDPs) were modeled using rational B-spline curve function, which gives $\alpha$ -NT2FRBsC(Fig. 5.2).

$$\overleftrightarrow{{}^{N}RBsC}^{\alpha}_{k,n}(t)=\frac{\sum_{i=1}^{n} w_i N_i^4(t)\,{}^{N}\overset{\leftrightarrow}{\vec{P}}_i^{\alpha}}{\sum_{r=1}^{n} w_r N_r^k(t)},\quad k=3;\ n=4;\ i=r=1,\dots,4.$$

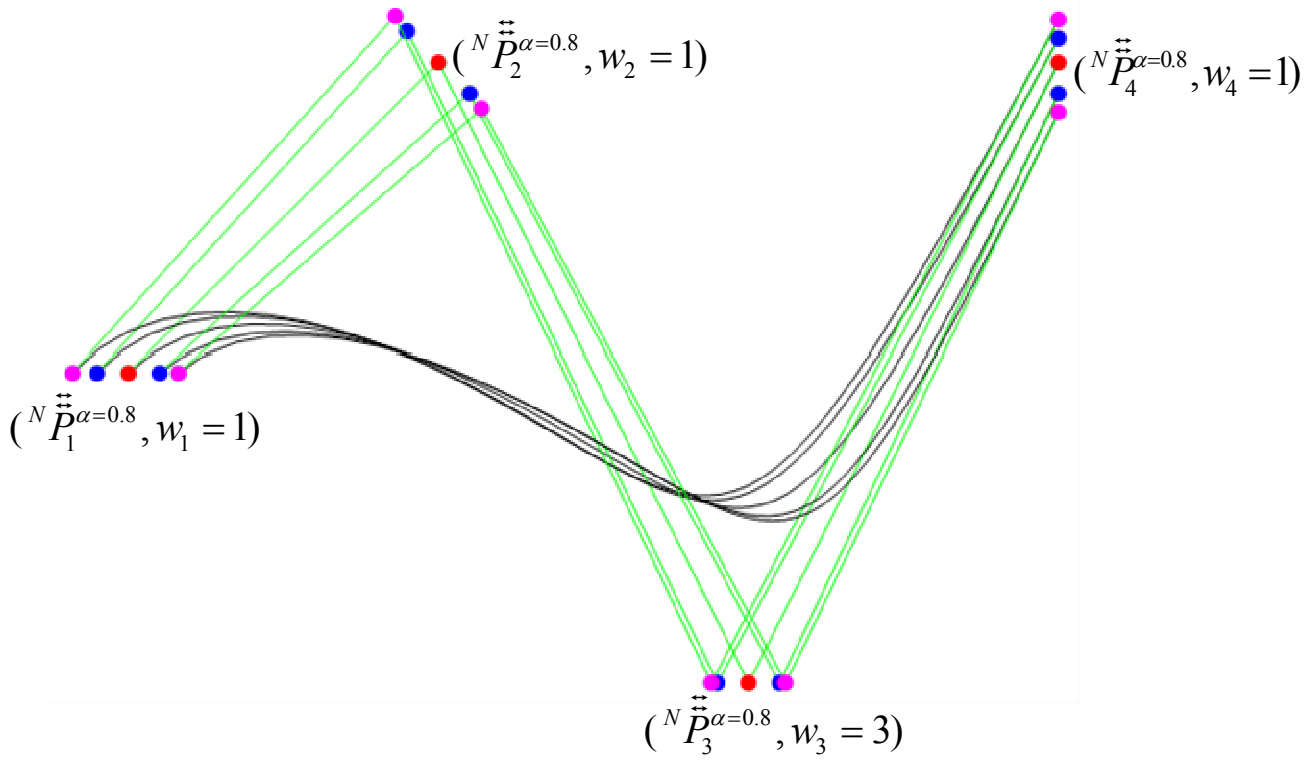

Figure 5.2. $\alpha$ -NT2FRBsC modeling with $w_{i=1,\dots,4}=(1,1,3,1)$.

Fig. 5.2 shows the modeling of $\alpha$ -NT2FRBsC in the cubic form with the value of $\alpha$ is 0.8.

**Step 3:** Type-reduction process of $\alpha$ -NT2FRBsC.

Based on Eq. 4.3 for $\alpha_{LMF} < \alpha_i < \alpha_{UMF}$, then

$$\begin{aligned}{}^{N}\bar{\vec{P}}^{\alpha}_{i=0,\dots,3} &= \left\langle \frac{1}{2}\sum_{i=1,\dots,4}\left\langle \overleftarrow{\vec{P}}_i^{\alpha\leftarrow} + \vec{P}_i^{\alpha\leftarrow} + 0 \right\rangle, P_i, \frac{1}{2}\sum_{i=1,\dots,4}\left\langle 0 + \vec{P}_i^{\alpha\rightarrow} + \overrightarrow{\vec{P}}_i^{\alpha\rightarrow} \right\rangle \right\rangle_{i=1,\dots,4} \\ &= \left\langle \bar{\vec{P}}_i^{\alpha\leftarrow}, P_i, \bar{\vec{P}}_i^{\alpha\rightarrow} \right\rangle.\end{aligned}$$

Then, the type-reduction of $\alpha$ -NT2FRBsC(TR- $\alpha$ -NT2FRBsC) model is,

$$\overleftrightarrow{{}^{N}RBsC}^{\alpha}_{k,n}(t)=\frac{\sum_{i=1}^{n} w_i N_i^k(t)\,{}^{N}\bar{\vec{P}}_i^{\alpha}}{\sum_{r=1}^{n} w_r N_r^k(t)},\quad k=3;\ n=4;\ i=r=1,\dots,4.$$

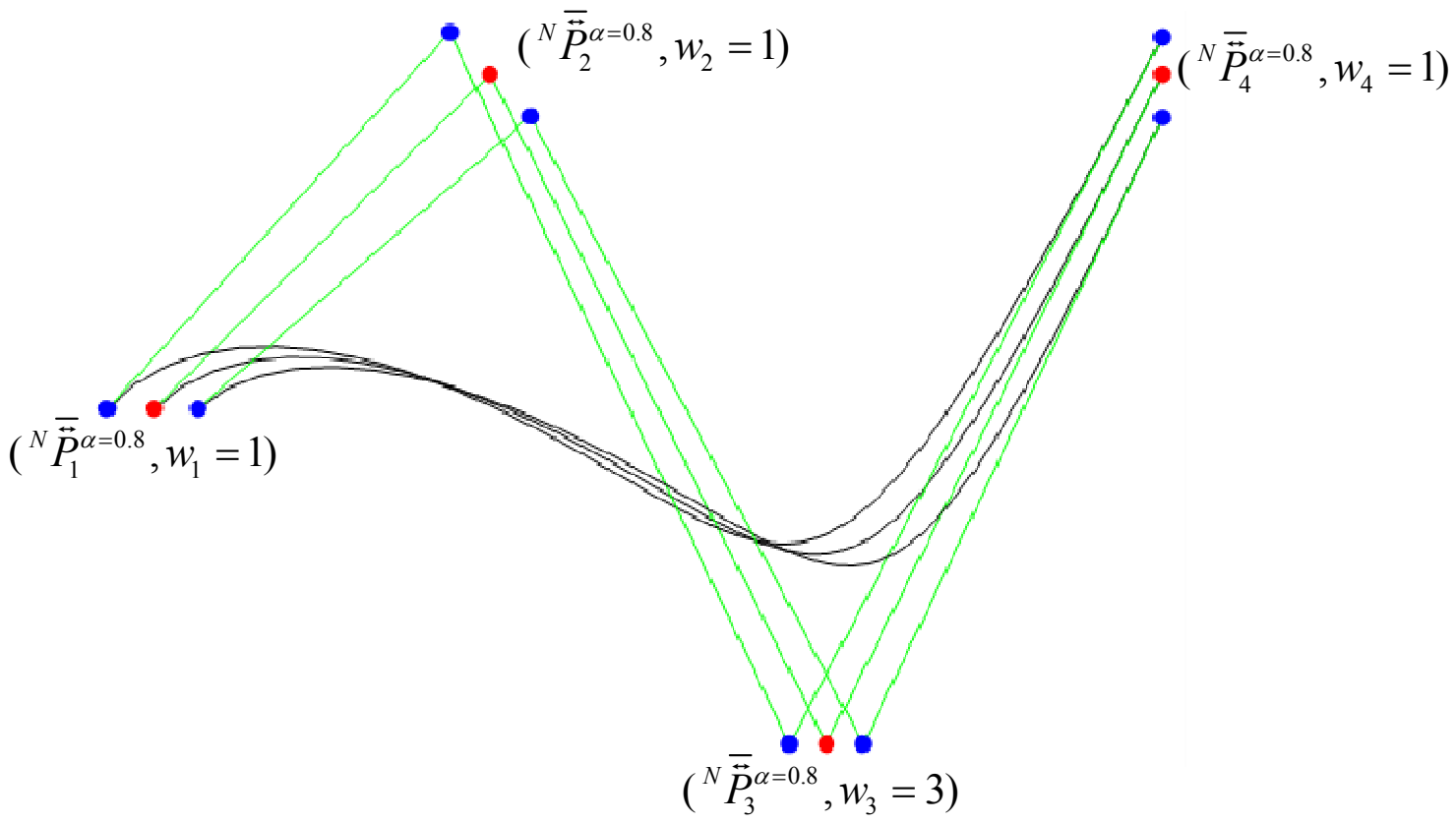

Figure 5.3. TR-$\alpha$ -NT2FRBsC modeling with $w_{i=1,\ldots,4} = (1,1,3,1)$.

Fig. 5.3 shows that the modeling of TR-$\alpha$ -T2FRBsC. The TR-$\alpha$ -T2FRBsC also becomes the type-1 fuzzy curve form which allows us to defuzzify the TR-$\alpha$ -T2FRBsC to become crisp T2FRBs solution curve(D-TR-$\alpha$ -T2FRBsC). The method used to defuzzify the TR-$\alpha$ -T2FRBsC is the centroid min method [2-4,34,39,37].

**Step 4:** Defuzzification of TR-$\alpha$ -T2FRBsC.

$$
\begin{aligned}
{}^{N}\overline{\overline{P}}_{i=1,\ldots,4}^{\alpha} &= \frac{1}{3}\sum_{i=0,\ldots,3}\left\langle \overleftrightarrow{\overline{P}}_i^{\alpha\leftarrow} + P_i + \overleftrightarrow{\overline{P}}_i^{\alpha\rightarrow} \right\rangle_{i=1,\ldots,4} \\
&= \overline{\overline{P}}_{i=1,\ldots,4}^{\alpha}.
\end{aligned}
$$

Then, the crisp T2FRBs solution curve model is given by Fig. 5.4.

$$
\overline{\overline{{}^{N}RBsC}}^{\alpha}_{k,n}(t) = \frac{\sum_{i=1}^{n} w_i N_i^k(t)\, {}^{N}\overline{\overline{P}}_i^{\alpha}}{\sum_{r=1}^{n} w_r N_r^k(t)}, \quad k = 3;\ n = 4;\ i = r = 1,\ldots,4.
$$

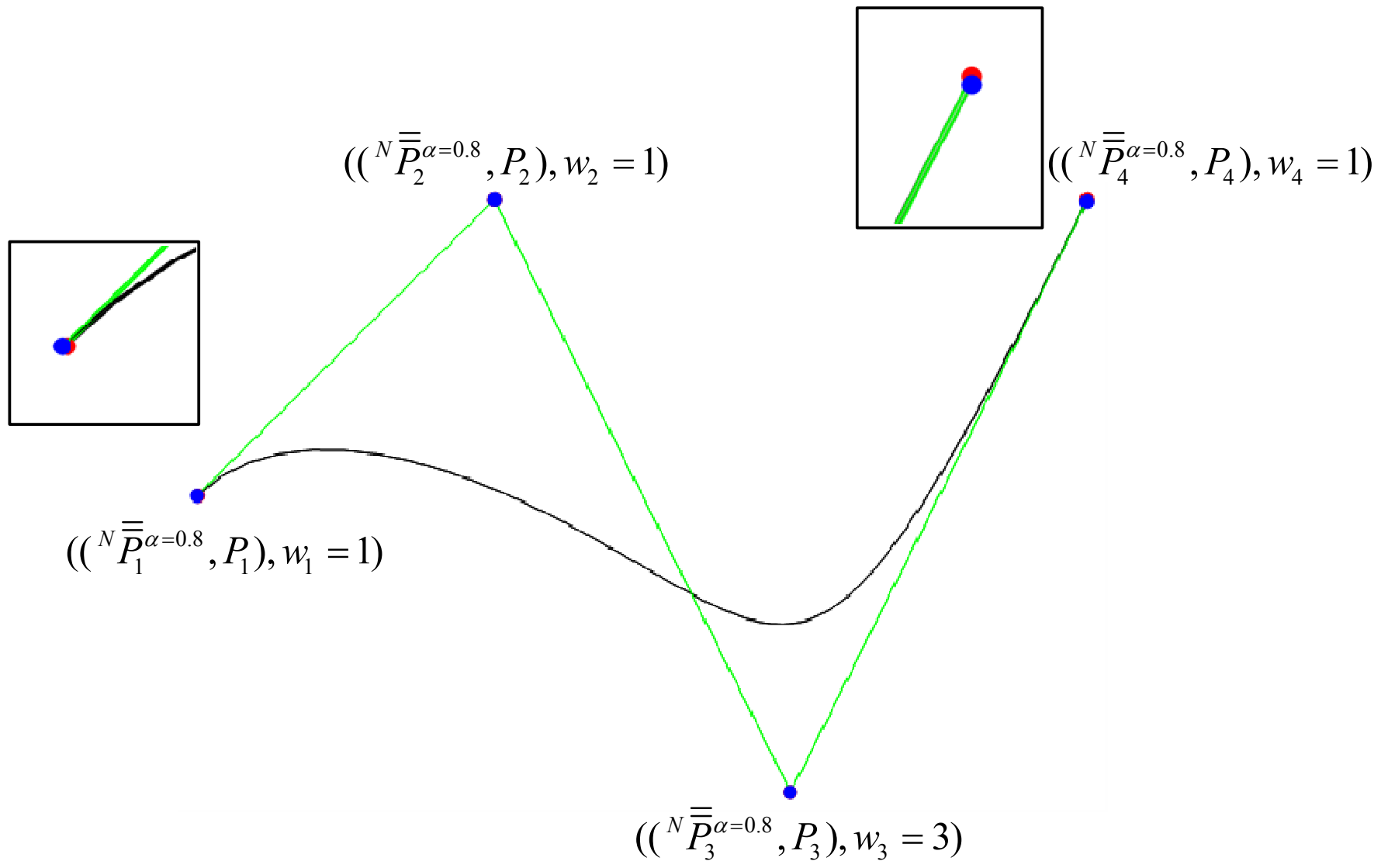

Figure 5.4. Crisp NT2FRBs solution curve, $\overline{\overline{{}^{N}RBsC_{k,n}}}^{\alpha}(t)$ marked by blue points together with crisp rational B-pline curve marked by red points.

Fig. 5.4 shows that the modeling of $\overline{\overline{{}^{N}RBsC_{k,n}}}^{\alpha}(t)$ along with the crisp rational B-spline curve. There are some errors between the crisp T2FDPs solution and data points. This is happened because the T2FDPs are not equally the same, meaning that the length between the left T2FDPs and crisp control points are not equal throughout the length between the right T2FDPs and crisp control points.

## 6 Conclusion

The construction of T2FDPs to deal with complex uncertainty data set had been constructed by using the fundamental of T2FST and being modeled by using rational B-spline curve function. The T2FDPs in the normal form(NT2FDPs) was defined, where this definition gives us advantage in dealing the variety of T2FDPs. This method can be applied in dealing with complex data uncertainty, which exists in real-life application. Currently, this method is extended for surface design by using tensor product technique for surface modeling.

## Acknowledgement

The authors would like to thank Research Management and Innovation Centre (RMIC) of Universiti Malaysia Terengganu and Ministry of Higher Education (MOHE) Malaysia for funding(FRGS, vot59244) and providing the facilities to carry out this research.